\documentstyle[twoside]{article}

\catcode`\@=11
\long\def\@makefntext#1{
\protect\noindent \hbox to 3.2pt {\hskip-.9pt  
$^{{\eightrm\@thefnmark}}$\hfil}#1\hfill}		

\def\@makefnmark{\hbox to 0pt{$^{\@thefnmark}$\hss}}	
	
\def\ps@myheadings{\let\@mkboth\@gobbletwo
\def\@oddhead{\hbox{}
\rightmark\hfil\eightrm\thepage}   
\def\@oddfoot{}\def\@evenhead{\eightrm\thepage\hfil
\leftmark\hbox{}}\def\@evenfoot{}
\def\sectionmark##1{}\def\subsectionmark##1{}}

\oddsidemargin=\evensidemargin
\newcounter{sectionc}\newcounter{subsectionc}\newcounter{subsubsectionc}
\renewcommand{\section}[1] {\vspace{12pt}\addtocounter{sectionc}{1} 
\setcounter{subsectionc}{0}\setcounter{subsubsectionc}{0}\noindent 
	{\tenbf\thesectionc. #1}\par\vspace{5pt}}
\renewcommand{\subsection}[1] {\vspace{12pt}\addtocounter{subsectionc}{1} 
	\setcounter{subsubsectionc}{0}\noindent 
	{\bf\thesectionc.\thesubsectionc. {\kern1pt \bfit #1}}\par\vspace{5pt}}
\renewcommand{\subsubsection}[1] {\vspace{12pt}\addtocounter{subsubsectionc}{1}
	\noindent{\tenrm\thesectionc.\thesubsectionc.\thesubsubsectionc.
	{\kern1pt \tenit #1}}\par\vspace{5pt}}

\newcounter{appendixc}
\newcounter{subappendixc}[appendixc]
\newcounter{subsubappendixc}[subappendixc]
\renewcommand{\thesubappendixc}{\Alph{appendixc}.\arabic{subappendixc}}
\renewcommand{\thesubsubappendixc}
	{\Alph{appendixc}.\arabic{subappendixc}.\arabic{subsubappendixc}}

\renewcommand{\appendix}[1] {\vspace{12pt}
        \refstepcounter{appendixc}
        \setcounter{figure}{0}
        \setcounter{table}{0}
        \setcounter{lemma}{0}
        \setcounter{theorem}{0}
        \setcounter{corollary}{0}
        \setcounter{definition}{0}
        \setcounter{equation}{0}
        \renewcommand{\thefigure}{\Alph{appendixc}.\arabic{figure}}
        \renewcommand{\thetable}{\Alph{appendixc}.\arabic{table}}
        \renewcommand{\theappendixc}{\Alph{appendixc}}
        \renewcommand{\thelemma}{\Alph{appendixc}.\arabic{lemma}}
        \renewcommand{\thetheorem}{\Alph{appendixc}.\arabic{theorem}}
        \renewcommand{\thedefinition}{\Alph{appendixc}.\arabic{definition}}
        \renewcommand{\thecorollary}{\Alph{appendixc}.\arabic{corollary}}
        \renewcommand{\theequation}{\Alph{appendixc}.\arabic{equation}}
        \noindent{\tenbf Appendix \theappendixc #1}\par\vspace{5pt}}
\newcommand{\subappendix}[1] {\vspace{12pt}
        \refstepcounter{subappendixc}
        \noindent{\bf Appendix \thesubappendixc. {\kern1pt \bfit #1}}
	\par\vspace{5pt}}
\newcommand{\subsubappendix}[1] {\vspace{12pt}
        \refstepcounter{subsubappendixc}
        \noindent{\rm Appendix \thesubsubappendixc. {\kern1pt \tenit #1}}
	\par\vspace{5pt}}

\newcommand{\textlineskip}{\baselineskip=13pt}
\newcommand{\smalllineskip}{\baselineskip=10pt}

\def\eightcirc{
\begin{picture}(0,0)
\put(4.4,1.8){\circle{6.5}}
\end{picture}}
\def\eightcopyright{\eightcirc\kern2.7pt\hbox{\eightrm c}} 

\newcommand{\copyrightheading}[1]
	{\vspace*{-2.5cm}\smalllineskip{\flushleft
	{\footnotesize International Journal of Modern Physics A, #1}\\
	{\footnotesize $\eightcopyright$\, World Scientific Publishing
	 Company}\\
	 }}

\def\abstracts#1#2#3{{
	\centering{\begin{minipage}{4.5in}\baselineskip=10pt\footnotesize
	\parindent=0pt #1\par 
	\parindent=15pt #2\par
	\parindent=15pt #3
	\end{minipage}}\par}} 

\renewenvironment{thebibliography}[1]
	{\frenchspacing
	 \ninerm\baselineskip=11pt
	 \begin{list}{\arabic{enumi}.}
	{\usecounter{enumi}\setlength{\parsep}{0pt}
	 \setlength{\leftmargin 12.7pt}{\rightmargin 0pt} 
	 \setlength{\itemsep}{0pt} \settowidth
	{\labelwidth}{#1.}\sloppy}}{\end{list}}

\newcounter{itemlistc}
\newcounter{romanlistc}
\newcounter{alphlistc}
\newcounter{arabiclistc}

\newcommand{\fcaption}[1]{
        \refstepcounter{figure}
        \setbox\@tempboxa = \hbox{\footnotesize Fig.~\thefigure. #1}
        \ifdim \wd\@tempboxa > 5in
           {\begin{center}
        \parbox{5in}{\footnotesize\smalllineskip Fig.~\thefigure. #1}
            \end{center}}
        \else
             {\begin{center}
             {\footnotesize Fig.~\thefigure. #1}
              \end{center}}
        \fi}

\newcommand{\tcaption}[1]{
        \refstepcounter{table}
        \setbox\@tempboxa = \hbox{\footnotesize Table~\thetable. #1}
        \ifdim \wd\@tempboxa > 5in
           {\begin{center}
        \parbox{5in}{\footnotesize\smalllineskip Table~\thetable. #1}
            \end{center}}
        \else
             {\begin{center}
             {\footnotesize Table~\thetable. #1}
              \end{center}}
        \fi}

\def\@citex[#1]#2{\if@filesw\immediate\write\@auxout
	{\string\citation{#2}}\fi
\def\@citea{}\@cite{\@for\@citeb:=#2\do
	{\@citea\def\@citea{,}\@ifundefined
	{b@\@citeb}{{\bf ?}\@warning
	{Citation `\@citeb' on page \thepage \space undefined}}
	{\csname b@\@citeb\endcsname}}}{#1}}

\newif\if@cghi
\def\cite{\@cghitrue\@ifnextchar [{\@tempswatrue
	\@citex}{\@tempswafalse\@citex[]}}
\def\citelow{\@cghifalse\@ifnextchar [{\@tempswatrue
	\@citex}{\@tempswafalse\@citex[]}}
\def\@cite#1#2{{$\null^{#1}$\if@tempswa\typeout
	{IJCGA warning: optional citation argument 
	ignored: `#2'} \fi}}

\def\pmb#1{\setbox0=\hbox{#1}
	\kern-.025em\copy0\kern-\wd0
	\kern.05em\copy0\kern-\wd0
	\kern-.025em\raise.0433em\box0}

\def\fnt#1#2{\footnotetext{\kern-.3em
	{$^{\mbox{\scriptsize #1}}$}{#2}}}

\def\fpage#1{\begingroup
\voffset=.3in
\thispagestyle{empty}\begin{table}[b]\centerline{\footnotesize #1}
	\end{table}\endgroup}

\def\runninghead#1#2{\pagestyle{myheadings}
\markboth{{\protect\footnotesize\it{\quad #1}}\hfill}
{\hfill{\protect\footnotesize\it{#2\quad}}}}
\font\tenrm=cmr10
\font\tenit=cmti10 
\font\tenbf=cmbx10
\font\bfit=cmbxti10 at 10pt
\font\ninerm=cmr9

\font\eightrm=cmr8

\def\qed{\hbox{${\vcenter{\vbox{			
   \hrule height 0.4pt\hbox{\vrule width 0.4pt height 6pt
   \kern5pt\vrule width 0.4pt}\hrule height 0.4pt}}}$}}

\begin{document}

\runninghead
{Economic Model of Neutrino Masses and Mixings}
{Economic Model of Neutrino Masses and Mixings}

\normalsize\textlineskip
\thispagestyle{empty}
\setcounter{page}{1}

\copyrightheading{}			

\vspace*{0.88truein}

\fpage{1}
\centerline{\bf ECONOMIC MODEL FOR NEUTRINO MASSES AND MIXINGS}
\vspace*{0.035truein}
\centerline{\footnotesize PAUL H. FRAMPTON}
\vspace*{0.035truein}
\centerline{\footnotesize\it 
Institute of Field Physics,
Department of Physics and Astronomy,}
\vspace*{0.035truein}
\centerline{\footnotesize\it
University of North Carolina,
Chapel Hill, NC 27599-3255}
\baselineskip=10pt
\vspace*{0.225truein}
\abstracts{Working in the framework of three chiral neutrinos
with Majorana masses, we investigate a scenario 
where the
neutrino mass matrix is strictly
off-diagonal in the flavor basis,  with all
its diagonal entries precisely zero.
}{}{}


\vspace*{1pt}\textlineskip	
\vspace*{0.5pt}
\bigskip
\bigskip
\noindent

This talk is based on a publication with Glashow\cite{FG} to which I
refer for more details than can be accommodated in this write-up.
The minimal standard model involves three chiral neutrino states, but
it does not admit renormalizable interactions that
can generate neutrino masses. Nevertheless, experimental
evidence suggests that
both solar and atmospheric neutrinos display flavor oscillations,
and  hence that neutrinos do have mass.
Two very different neutrino squared-mass differences are required to fit
the data:
\begin{equation}
10^{-11}{\rm eV^2} \le  \Delta_s \le 10^{-5}{\rm eV^2}
~~~ {\rm and} ~~~ \Delta_a\simeq 10^{-3}{\rm eV^2}
\label{edata}
\end{equation}
where the neutrino masses $m_i$ are ordered  such that:
\[
\Delta_s \equiv | m_2^2-m_1^2 | ~~~{\rm and}~~~ \Delta_a \equiv \vert m_3^2-m_2^2
\vert \simeq \vert m_3^2-m_1^2 \vert
\]
and the subscripts $s$ and $a$ pertain to solar and atmospheric
oscillations respectively.
 The large
uncertainty in $\Delta_s$ reflects  the several potential  explanations
of the observed solar neutrino flux: in terms of vacuum oscillations or
large-angle  or  small-angle MSW solutions,  but in
every case the two independent squared-mass differences must be widely
spaced with
\[
r\equiv \Delta_s/\Delta_a < 10^{-2}
\]

Solar neutrinos
may exhibit an energy-independent time-averaged suppression due to
$ \Delta_a $, as well as energy-dependent oscillations depending on
$\Delta_s/E$. Atmospheric neutrinos may exhibit oscillations due to
$\Delta_a$, but they
are almost entirely unaffected by $\Delta_s$. It is convenient
to define
neutrino mixing angles as follows:
\[
\left( \begin{array}{c}\nu_e \\\nu_{\mu} \\ \nu_{\tau} \end{array} \right)
=
\left( \begin{array}{ccc}c_2c_3 & c_2s_3 & s_2e^{-i\delta} \\
+c_1s_3 +s_1s_2c_3e^{i\delta}& -c_1c_3-s_1s_2s_3e^{i\delta}&-s_1c_2 \\
+s_1s_3 -c_1s_2c_3e^{i\delta}& -s_1c_3-c_1s_2s_3e^{i\delta}&+c_1c_2
\end{array} \right)
\left( \begin{array}{c} \nu_1 \\ \nu_2 \\ \nu_3 \end{array} \right)
\]
with $s_i$ and $c_i$ standing for sines and cosines of $\theta_i$.
For neutrino masses satisfying Eq.(\ref{edata})
the vacuum survival  probability of solar neutrinos is\cite{GG}

\begin{equation}
P(\nu_e\rightarrow\nu_e)\big\vert_s \simeq 1-{\sin^2{2\theta_2}\over2}
- \cos^4{\theta_2}\sin^2{2\theta_3}sin^2{(\Delta_s R_s/4E)}
\label{eproba}
\end{equation}
whereas the transition probabilities  of atmospheric neutrinos are:

\begin{eqnarray}
P(\nu_\mu  \rightarrow\nu_\tau)\big\vert_a  & \simeq &
\sin^2{2\theta_1}\cos^4{\theta_2}\,\sin^2{(\Delta_a R_a/4E)} \nonumber  \\
 P(\nu_e \leftrightarrow\nu_\mu)\big\vert_a & \simeq &
\sin^2{2\theta_2}\sin^2{\theta_1}\,\sin^2{(\Delta_aR_a/4E)} \nonumber \\
 P(\nu_e \rightarrow\nu_\tau)\big\vert_a  & \simeq &
\sin^2{2\theta_2}\cos^2{\theta_1}\,\sin^2{(\Delta_aR_a/4E)} \nonumber \\
\end{eqnarray}

None of these probabilities depend on $\delta$, the measure of CP violation
in the lepton sector.

Let us turn to the origin of neutrino masses.
Among the many
renormalizable and gauge-invariant extensions of the standard model
that can do the trick is the introduction  of a charged singlet meson $f^+$
coupled antisymmetrically to pairs of lepton doublets {\it and\/} (also
antisymmetrically) to a pair of Higgs doublets.  This
simple mechanism was first proposed by Zee\cite{zee}
and
results (at one loop) in a Majorana mass matrix
in the flavor basis ($e,\mu,\tau$)
 of a  special form:
\begin{equation}
{\cal M}= \left( \begin{array}{ccc} 0 & m_{e\mu} & m_{e\tau} \\
 m_{e\mu} & 0 & m_{\mu\tau} \\
                   m_{e\tau} & m_{\mu\tau} & 0 \end{array} \right)
\label{ansatz}
\end{equation}

\noindent
Related  discussions of Eq.(\ref{ansatz})
appear elsewhere\cite{J,ST}
The present work is essentially a continuation of \cite{J}.

The sum of the neutrino masses (the eigenvalues of $\cal M$)
vanishes:
\begin{equation}
 m_1+m_2+m_3=0
\label{etrace}
\end{equation}
An important result  emerges when the squared-mass hierarchy
Eqs.(\ref{edata}) is taken into account along with Eq.(\ref{etrace}).
In the limit $r\rightarrow 0$, two of the squared masses must be equal.
There are two possibilities.
 In case A, we have
$m_1+m_2 = 0$ and $m_3= 0$. This case arises
iff at least one of
the three parameters in $\cal M$  vanishes.
In case B, we have $m_1= m_2$  and  $m_3= -2m_2>0$. This case
arises iff the three parameters in $\cal M$ are equal to one another.
Of course, $r$ is small but it does not vanish:  in neither case can
the above relations among neutrino masses be strictly satisfied.
But they must be nearly
satisfied. Consequently  we  may deduce
certain approximate but useful
restrictions on the permissable values of the
neutrino mixing angles $\theta_i$. Prior to examining these restrictions,
we note that Eqs.(\ref{etrace}) and (\ref{edata})
exclude the possibility that the three neutrinos are nearly degenerate in
mass.  

In \cite{FG} these cases are analysed carefully and the best
version is an example of Case A with
$m_{\mu\tau} = 0$
and leads to conservation of $L_e-L_\mu-L_\tau$.
 For this subcase, we obtain $\sin{\theta_2}=0$ and
$\theta_3=\pi/4$. We see from Eq.(\ref{eproba})  that solar
neutrino oscillations are maximal:
\begin{equation}
P(\nu_e\rightarrow\nu_e)\big\vert_s=
1-\sin^2{(\Delta_s R_s/4E)}
\label{esolarosc}
\end{equation}
Moreover, we see from Eq.(\ref{eproba})
that atmospheric $\nu_\mu$'s oscillate exclusively into $\nu_\tau$'s with
the unconstrained mixing angle $\theta_1$:
\begin{equation}
P(\nu_\mu\rightarrow \nu_\tau)\big\vert_a
 =\sin^2{2\theta_1}\,\sin^2{(\Delta_a
R_a/4E)}\\
P(\nu_\mu \leftrightarrow \nu_e)\big\vert_a = 0
\quad\quad P(\nu_e\rightarrow \nu_\tau)\big\vert_a=0\
\label{eatmososc}
\end{equation}

This implementation is compatible with experiment:
It predicts maximal energy-independent
solar oscillations
and it remains consistent with the solar data if chlorine 
results are dropped.

\bigskip 
\bigskip

The advantage of the model is its economy: 
no additional chiral fermions are present and so, 
unlike other extensions\cite{catchall} of the standard model 
anomaly cancellation does not lead to any 
new restrictions, and the number of added parameters is small.

\end{document}